\documentstyle[sprocl]{article}
\input epsf
\def\nrcpt{NR\raise.4ex\hbox{$\chi$}PT\ }

\def\ltap{\ \raise.3ex\hbox{$<$\kern-.75em\lower1ex\hbox{$\sim$}}\ }
\def\gtap{\ \raise.3ex\hbox{$>$\kern-.75em\lower1ex\hbox{$\sim$}}\ }

\def\CA{{\cal A}}

\def\CL{{\cal L}}

\def\pds{{\it PDS}\ }
\def\ms{MS}

\def\bfq{{\bf q}}

\def\bfv{{\bf v}}
\def\bfp{{\bf p}}

\def\frac#1#2{{\textstyle{#1\over#2}}}
\def\darr#1{\raise1.5ex\hbox{$\leftrightarrow$}\mkern-16.5mu #1}
\def\){\right)}
\def\({\left(}
\def\]{\right]}
\def\[{\left[}
\def\si{{}^1\kern-.14em S_0}
\def\siii{{}^3\kern-.14em S_1}
\def\diii{{}^3\kern-.14em D_1}

\def\CA{{\cal A}}

\newcommand{\eqn}[1]{\label{eq:#1}}
\newcommand{\refeq}[1]{(\ref{eq:#1})}
\newcommand{\eq}{eq.~\refeq}

\newcommand{\eqsii}[2]{eqs.~(\ref{eq:#1}, \ref{eq:#2})}

\newcommand{\beq}{\begin{eqnarray}}% can be used as {equation} or {eqnarray}
\newcommand{\eeq}{\end{eqnarray}}
\def\Journal#1#2#3#4{{#1} {\bf #2}, #3 (#4)}
\def\NPB{{\em Nucl. Phys.} B}

\def\PLB{{\em Phys. Lett.}  B}

\def\PRC{{\em Phys. Rev.} C}

\begin{document}
\title{Power Counting and the Renormalization Group for an Effective
Description of $NN$ Scattering\footnote{Talk presented at the Joint Caltech/INT
Workshop on Nuclear Physics with Effective Field Theory, February 1998. Report
No. DOE/ER/40561-6-INT98
 }}
\author{David B. Kaplan}
\address{Institute for Nuclear Theory 351550, University of Washington,\\
Seattle WA, 98195-1550
\\}
\maketitle

\abstracts{
I outline the power counting scheme recently introduced by  M. Savage, M. Wise
and
myself for the effective field theory treatment of $NN$ scattering.  It is
particularly useful
for describing systems with a large scattering length, and differs from
Weinberg's power counting.
A renormalization group analysis plays a big role in determining the order of a
given operator. Pions are ignored in this discussion;  how to incorporate them
is discussed in M. Savage's talk~\cite{Martin}.}
%%%%%%%%%%%%%%%%%%%%%%%%%
\section{Introduction}
%%%%%%%%%%%%%%%%%%%%%%%%%

The purpose of this talk is to outline an effective field theory expansion for
$NN$ scattering
recently developed in Ref.~\cite{KSW} with my collaborators \footnote{These notes
contain somewhat more
than my talk did, not out of
revisionist tendencies, but  because discussions at the workshop spurred us to
compute the
electromagnetic form factors of the deuteron~\cite{Deut}, and in the
process I have expanded my
understanding of the subject.}.
When pions are included (Martin Savage's talk in this volume), this expansion
differs
from the original approach advocated by Weinberg~\cite{Weinberg}  and
implemented or discussed since
Refs.~\cite{Bira,Park,KSWa,CoKoM,Cohen,Lepage,Adhik,RBMa,Gegelia,Steele}.
 The advantage we claim for our
scheme is that it allows analytic calculations of two-nucleon processes in a
controlled expansion.
As a result, the calculations of low energy  $NN$ processes are on the
same footing as meson interactions in chiral perturbation theory ($\chi PT$),
or meson-baryon
interactions in heavy-baryon chiral perturbation theory ($HB\chi PT$).  An
obvious advantage of
this approach is that it is much easier to extract physics from an analytical
formula than from
a computer simulation;  for example, when one computes a Green function in
terms of external
momenta from Feynman diagrams, on has the whole analytic structure over a wide
kinematic range
and can analyze inelastic processes by examining the cuts, etc. Furthermore,
one can better dissect
the result and understand the relative importance from the various
contributions.  Finally, as we
demonstrate explicitly in Ref.~\cite{Deut},  by being able to perform analytic
calculations one can
retire the old spectra of ``off-shell ambiguity'', showing it to be hardly
worth
the fear and loathing that it usually inspires. No $S$-matrix element depends
on off-shell quantities!

One might think that in choosing to do an effective field theory calculation of
$NN$ interactions
analytically rather than numerically, one is sacrificing accuracy for
elegance.  In fact the power
counting suggests this is not the case...Weinberg's proposal to expand the
potential, and then resum
its effects to all orders via the Lippmann-Schwinger equation
to get the $S$-matrix is not  more accurate than simply
 expanding the $S$-matrix directly.  Furthermore, it does not even constitute
an expansion
unless one  chooses the cutoff in the theory with some care. Therefore I see no
virtue and some
disadvantages with performing an effective field theory expansion of the
potential, and then solving
the Lippmann-Schwinger equation numerically.  I believe that nuclear physics
could well profit from
abandoning such traditional tools as potentials and wave functions.

Much emphasis has been placed on regulating and renormalizing the effective
theory of $NN$
interactions.  In fact, I view this workshop as the opportunity to put this
issue behind us.
The only reason to focus on these issues is to develop a consistent power
counting scheme;
once such a scheme is developed and followed consistently, results  will not be
scheme dependent, or else one is working with a model of $NN$ interactions,
as opposed to an effective field theory.
So, while I emphasize in this talk the renormalization
scheme used, the purpose is to develop the tools so that we can do physics, and
not out of a love for formalism.

%%%%%%%%%%%%%%%%%%%%%%%%%%%%%%%%%%%%%%%%%%%%%%%%%%%%%%%%%%%%%%%%%%%%%%%%%%%%%%
\section{Effective field theory for nonrelativistic scattering: a toy example}
\label{sec:2}
%%%%%%%%%%%%%%%%%%%%%%%%%%%%%%%%%%%%%%%%%%%%%%%%%%%%%%%%%%%%%%%%%%%%%%%%%%%%%%

Consider  a toy model of heavy spinless ``nucleons'' $\tilde N$
interacting via a Yukawa interaction characterized by a scale $\Lambda$. For
low energy
scattering we can
construct the effective field theory describing scattering at
 momenta $p\ll \Lambda$,  consisting entirely of
contact interactions in a derivative expansion.
 Since these local operators are singular, the formulation of the low energy
theory  necessarily introduces divergences that can be dealt with by
conventional regularization and renormalization procedures, so that the final
result is independent of a momentum cutoff. I will  show how to organize the
Feynman
graphs in the effective theory in a consistent power counting scheme so that
the
scattering amplitude can be expanded in powers of $p/\Lambda$. Since
the sizes of all the coupling constants in the effective theory depend on the
subtraction scheme used to render diagrams finite, the development of the power
counting scheme is intimately related to the renormalization procedure used.  I
will explain  why the \pds  subtraction scheme introduced in Ref.~\cite{KSW}\
is particularly
well suited for this problem.

 It should be no surprise that the effective field theory expansion for the
toy system is simply related to the conventional effective range expansion,
 and so the machinery of quantum field theory may appear to be heavy handed
and superfluous. Nevertheless,  the field theoretic language that I develop
in this section is readily extended to the realistic problem of interest:
nucleons  interacting via both short range interactions and long range pion
exchange (see M. Savage's talk, Ref.\cite{Martin}).  In the realistic problem,
effective field theory is not equivalent
to an effective range expansion, and is the only framework that can
consistently incorporate chiral
symmetry and relativistic effects without resorting to
phenomenological  models.

Assume that the spinless bosons $\tilde N$ are nonrelativistic with  mass $M$,
carry a conserved charge (``baryon
number''), and interact via the exchange of a meson $\phi$ with mass $\Lambda$
and coupling $g$.  At tree level, meson exchange gives rise to the Yukawa
interaction
\beq
V(r)= -{g^2\over 4\pi} {e^{-\Lambda r}\over r},
\eeq
and the Schr\"odinger equation for this system may be written as
\beq
\left[ -\nabla_x^2 + \eta {e^{-x}\over x} - {p^2\over \Lambda^2}\right]\Psi=0\
,\\
 \vec x\equiv \Lambda \vec r, \qquad \eta\equiv{g^2 M\over 4\pi \Lambda},
\qquad p^2\equiv M E\ .
\eqn{sequ}
\eeq
Note that $p$ is the magnitude of the 3-momentum carried by each $\tilde N$
particle in the center of mass frame.
Evidently there are two options for a perturbative solution for the $S$-matrix
for this system.  The first is an expansion in powers of $\eta$, the familiar
Born
expansion.
An alternative is to expand in powers of $p/\Lambda$, which is
the expansion parameter used in effective field theory.   An important feature
of the low energy expansion is that it can provide accurate results in terms
of a few phenomenological parameters even for nonperturbative $\eta$.  I will
assume throughout that
 $\eta\sim 1$, since this is
the regime we are interested in (a strongly coupled system without a plethora
of bound states).

The quantity that is natural to calculate in a field theory is the sum of
Feynman
graphs, which gives the amplitude $i\CA$, related to the $S$-matrix by
\beq
S= 1 + i {Mp\over 2\pi}\CA\ .
\eeq
For $S$-wave scattering, $\CA$ is related to the phase shift $\delta$
by
\beq
\CA = {4\pi \over M}{1\over p\cot\delta -i p}\ .
\eqn{amp}
\eeq
{}From quantum mechanics it is well known that it is not $\CA$, but rather the
quantity
$p\cot\delta$, which has a nice momentum
expansion for $p\ll\Lambda$ (the effective range expansion):
\beq
p\cot\delta = -{1\over a} + {1\over 2}\Lambda^2\sum_{n=0}^\infty {r}_n
\({p^2\over \Lambda^2}\)^{n+1}\ ,
\eqn{erexp}
\eeq
where $a$ is the scattering length, and
$r_0$ is the effective range.
So long as $\eta\sim 1$
the coefficients  $r_n$ are generally $O(1/\Lambda)$ for all $n$. However,  $
a$ can
take on any value, diverging as $\eta$ approaches one of the critical couplings
$\eta_k$ for which there is a boundstate at threshold. (The lowest critical
coupling
is found numerically to be $\eta_1=1.7$.) Therefore the radius of convergence
of a
momentum expansion of $\CA$ depends on the size
of the scattering length $a$. First I consider the situation where the
scattering length is of natural
size $|a|\sim 1/\Lambda$, and then I  discuss the case
$|a|\gg 1/\Lambda$, which is   relevant for realistic $NN$ scattering.

%%%%%%%%%%%%%%%%%%%%%%%%%%%%%%%%%%%%%%%%%%%%%%%%%%%%%%%%%%%%%%%%
\subsection{The momentum expansion for a scattering length of natural size}
\label{sec:2a}
%%%%%%%%%%%%%%%%%%%%%%%%%%%%%%%%%%%%%%%%%%%%%%%%%%%%%%%%%%%%%%%%

In the regime $|a|\sim  1/\Lambda$ and $|r_n|\sim  1/\Lambda$,
the amplitude $\CA$ has a simple momentum expansion in terms of
the low energy scattering data,
\beq
\CA = -{4\pi a \over M }\[ 1 - i a p +(ar_0/2-a^2) p^2 + O(p^3/\Lambda^3)\]\ ,
\eqn{aexp}
\eeq
which converges up to momenta $p\sim \Lambda$.   It is this expansion that we
wish to reproduce in an effective field theory.

The effective field theory of $\tilde N$ particles
interacting  through contact interactions has the following  Lagrangian:
\beq
 \CL_{eff}&=&
{{\tilde N}^\dagger} \left( i\partial_t + \nabla^2/2M\right) \tilde N
\nonumber\\ &&+(\mu/2)^{4-D}\left[C_0 ({{\tilde N}^\dagger} \tilde N)^2
+ {C_2\over 8} \[(\tilde N \tilde N)^\dagger( N {
\mathop\nabla^\leftrightarrow}{ }^2 \tilde N) +h.c\]
+...\right]\ ,
\eqn{yukeff}
\eeq
where
\beq
 {  \mathop\nabla^\leftrightarrow}^2 \equiv  {  \mathop\nabla^\leftarrow}^2 -2
{ \mathop\nabla^\leftarrow \cdot \mathop\nabla^\rightarrow}+ {
\mathop\nabla^\rightarrow}^2\ .
\eeq
The sum of Feynman diagrams computed in this theory gives us the
amplitude $\CA$.
As I will be using dimensional regularization for the loop integrals in this
theory,
the spacetime dimension is given by $D$
\footnote{Dimensional regularization is the preferred regularization scheme
as it preserves gauge symmetry and chiral symmetry,
as well as Lorentz invariance (or Galilean invariance, for nonrelativistic
systems). The advantage of the latter is that it makes the Feynman integrals
easier to perform, as one can shift the integration variable.}.
The ellipsis indicates higher derivative operators,
and $(\mu/2)$ is an
arbitrary mass scale introduced to allow the couplings $C_{2n}$
multiplying operators containing $\nabla^{2n}$  to have the same dimension for
any $D$.
 I focus on  the $s$-wave channel
(generalization to higher partial waves is straightforward), and assume
that $M$ is very large so that relativistic effects can be ignored. The form of
the
$C_2$ operator is fixed by Galilean invariance, which implies that when all
particle
momenta are boosted $\bfp \to \bfp + M\bfv$, the Lagrangian must remain
invariant.
There exists another two derivative operator for $p$-wave scattering which I
will not be discussing.

In general, the tree level $s$  partial wave amplitude in the center of mass
frame arising from $\CL_{eff}$ is
\beq
i \CA_{\rm tree}^{(cm)} = -i(\mu/2)^{4-D} \sum_{n=0}^\infty C_{2n}(\mu) p^{2n}\
,
\eqn{tree}
\eeq
where the coefficients $C_{2n}(\mu)$ are  the
couplings in the Lagrangian of operators with $2n$ gradients contributing to
$s$-wave scattering.  One may always
trade time derivatives for spatial gradients, using the equations of motion
when computing
$S$-matrix elements, and so we ignore such operators (see appendix~B in
Ref.\cite{Deut} for an explicit example)..

\begin{figure}[t]
\centerline{\epsfxsize=4.5 in \epsfbox{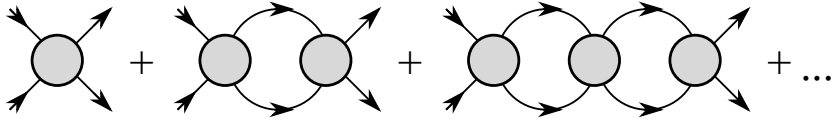}}
\noindent
\caption{\it The bubble chain arising from local operators. 
The vertex is given by the tree level amplitude, \eq{tree}.}
\label{bubbles}
\vskip .2in
\end{figure}
The loop integrals
one encounters in diagrams shown in Fig.~\ref{bubbles}
are of the form
\beq
\openup3\jot
I_n&\equiv& i(\mu/2)^{4-D} \int {{\rm d}^D q\over (2\pi)^D}\, {{\bf q}^{2n}
\over\( E/2 + q_0 -{{\bf q}^2\over 2M} + i\epsilon\)\( E/2 - q_0
-{{\bf q}^2\over 2M} + i\epsilon\)}
\nonumber\\
&=& (\mu/2)^{4-D} \int {{{\rm d}}^{(D-1)}{\bf  q}\over (2\pi)^{(D-1)}}\,
{\bf q}^{2n} \({1\over E  -{\bf q}^2/M + i\epsilon}\)
\nonumber\\
&=& -M (ME)^n (-ME-i\epsilon)^{(D-3)/2 } \Gamma\({3-D\over 2}\)
{(\mu/2)^{4-D}\over  (4\pi)^{(D-1)/2}}\ .
\eqn{loopi}
\eeq

In order to define the theory, one must specify a subtraction scheme; different
subtraction
schemes amount to a reshuffling between contributions from
the vertices and contributions from the the UV part of the
loop integration.  How does one choose a subtraction scheme that is useful?
I am considering the case $|a|, |r_n|\sim 1/\Lambda$, and wish to reproduce the
expansion of
the amplitude \eq{aexp}.  In order to do this via Feynman diagrams, it is
convenient if any Feynman
graph with a particular set of operators at the vertices only contributes to
the expansion of the
amplitude at a particular order.  Since the the expansion \eq{aexp} is a strict
Taylor expansion in $p$, it is
it is therefore very convenient if each Feynman graph gives one a simple
monomial in $p$.  Obviously,
this won't be true in a random subtraction scheme.
 A subtraction scheme that fulfills this criterion is the minimal subtraction
scheme ($\ms$) which
amounts to subtracting any
$1/(D-4)$ pole before taking the $D\to 4$ limit.
As the integral \eq{loopi} doesn't exhibit any such poles, the result in $\ms$
is simply
\beq
I_n^{\ms}=  (ME)^n \({M\over 4\pi}\)\sqrt{-ME-i\epsilon}=
-i\({M\over 4\pi}\) p^{2n+1}  \ .
\eqn{inval}
\eeq
Note the  nice feature of this scheme that the factors of $q$ inside the loop
get converted
to factors of $p$, the external momentum.  Similarly, a factor of the equations
of motion,
$i\partial_t +\nabla^2/2M$, acting on one of the internal legs at the vertex,
causes the
loop integral to vanish.
Therefore one can use the on-shell,
tree level amplitude \eq{tree} as the internal  vertices in loop diagrams;
summing the bubble diagrams in the center of mass frame gives
\begin{eqnarray}
\CA & = & -{  \sum C_{2n} p^{2n}
\over
1 + i(M p/4\pi) \sum C_{2n} p^{2n} }
\ .
\end{eqnarray}
 Since there are no poles at $D=4$ in the $\ms$ scheme, the coefficients
$C_{2n}$ are
independent of the subtraction
point $\mu$.
The power counting in the $\ms$ scheme is particularly simple, as promised:
\begin{enumerate}
\item{Each propagator counts as $1/p^2$;}
\item{Each loop integration $\int {\rm d}^4q$ counts as $p^5$ (since $q_0\sim
\bfq^2/2M$);}
\item{Each vertex $C_{2n}\nabla^{2n}$ contributes $p^{2n}$.}
\end{enumerate}
The amplitude may be expanded in powers of $p$ as
\beq
\CA=\sum_{n=0}^\infty \CA_n
\  ,\qquad
\CA_n\sim O(p^n)
\eqn{ampexpand}
\eeq
where the $\CA_n$ each arise from graphs with $L\le n$ loops and can be equated
to the low
energy scattering data \eq{aexp} in order to fit the $C_{2n}$ couplings.  In
particular,
$\CA_0$ arises from the tree graph with $C_0$ at the vertex; $\CA_1$ is given
by the
1-loop diagram with two $C_0$ vertices; $\CA_2$ is gets contributions from both
 the 2-loop diagram with
three $C_0$ vertices, as well as the tree diagram with one $C_2$ vertex, and so
forth.
Thus the first three terms are
\beq
\CA_0= -C_0\ ,\qquad
\CA_1= i C_0^2{Mp\over 4\pi}\ ,\qquad
\CA_2=  C_0^3\({Mp\over 4\pi}\)^2-C_2p^2
\ .
\eqn{effexp}
\eeq
Comparing \eqsii{aexp}{effexp} I find for the first two couplings of the
effective theory
\beq
C_0 = {4\pi a\over M}\ ,\qquad C_2 = C_0 {a r_0\over 2}
\ .
\eqn{cfit}
\eeq
In general, when the scattering length has natural size,
\beq
C_{2n} \sim {4\pi \over M\Lambda} {1\over \Lambda^{2n}}
\ .
\eeq
Note that the effective field theory calculation in this scheme
is completely perturbative even though $\eta\sim 1$ and
there may be a boundstate well below threshold.
The point is,
that when there are  no poles in $\CA$ in the region
$|p|\ltap\Lambda$, the amplitude is amenable
to a Taylor expansion in $p/\Lambda$ in that region;
with a suitable subtraction scheme ($\ms$), this
Taylor expansion can correspond to a perturbative sum of Feynman graphs.

%%%%%%%%%%%%%%%%%%%%%%%%%%%%%%%%%%%%%%%%%%%%%%%%%%%%%%%%%%%%
\subsection{The momentum expansion for large scattering length}
\label{sec:2b}
%%%%%%%%%%%%%%%%%%%%%%%%%%%%%%%%%%%%%%%%%%%%%%%%%%%%%%%%%%%%

Now consider the case $|a|\gg 1/\Lambda$, $|r_n|\sim 1/\Lambda$, which
is of relevance to realistic $NN$ scattering.
For a nonperturbative interaction ($\eta\sim 1$) with
a boundstate near threshold, the  expansion of $\CA$ in powers of $p$
is of little practical value, as it breaks down for momenta $p\gtap 1/|a|$,
far below $\Lambda$.
In the above effective theory, this occurs because the couplings $C_{2n}$
are anomalously large, $C_{2n}\sim 4\pi a^{n+1} / M\Lambda^n$.
However, the problem is
not with the effective field theory method,
but rather with the subtraction scheme chosen.

Instead of reproducing the expansion of the amplitude shown in \eq{aexp}, one
needs
to expand in powers of $p/\Lambda$ while retaining $ap$ to all orders:
\beq
\CA = -{4\pi\over M}{1\over (1/a + i p)}\[ 1 + {r_0/2 \over (1/a + ip)}p^2 +
{(r_0/2)^2\over (1/a + ip)^2} p^4 + {(r_1/2\Lambda^2)\over (1/a + ip) } p^4
+\ldots\]
\eqn{aexp2}
\eeq
Note that  for $p>1/|a|$ the terms  in this expansion scale as $\{p^{-1},
p^0,p^1,\ldots\}$.
Therefore, the expansion in the effective theory should take the form
\beq
\CA=\sum_{n=-1}^\infty \CA_n
\ ,\qquad
\CA_n\sim O(p^n)
\eqn{ampbiga}
\eeq
beginning at $n=-1$ instead of $n=0$, as in the expansion \eq{ampexpand}.
Comparing with \eq{aexp2}, we see that
\beq
\CA_{-1} &=&  -{4\pi\over M}{1\over (1/a + i p)}\ ,\nonumber\\
\CA_{0} &=&  -{4\pi\over M}{r_0p^2/2\over (1/a + i p)^2}\ ,
\eqn{ami}
\eeq
and so forth.
Again, the task is to compute the $\CA_n$ in the
effective theory, and  equate to the appropriate expression
 above, thereby fixing the
$C_{2n}$ coefficients.  As before, the goal is actually more ambitious:  each
particular
graph contributing to $\CA_n$ should be $O(p^n)$, so that the power counting is
transparent.

 As any single diagram in the effective theory
is proportional to positive powers of $p$, computing the leading term
$\CA_{-1}$ must involve
summing an infinite set of diagrams. It is easy to see that the leading term
$\CA_{-1}$
can be reproduced by the sum of bubble diagrams with $C_0$ vertices~\cite{Weinberg},
which yields in the $\ms$ scheme
\beq
{\cal A}_{-1} = { -C_0\over \left[1 + {C_0 M\over 4\pi} ip\right]}\ .
\eeq
Comparing this with \eq{ami} gives $C_0=4\pi a/M$, as in the
previous section.  However, there is no  expansion parameter that justifies
this summation:
each individual graph in the bubble sum goes as $C_0 (C_0 M p)^L\sim(4\pi
a/M)(i a p)^L$, where $L$ is the number
of loops. Therefore each graph in the bubble sum is bigger than the preceding
one, for $|ap|>1$,
while they sum up to something small.

This is an unpleasant situation for an effective field theory;  it is important
to have an
expansion parameter so that one can  identify the order of any particular
graph, and sum the
graphs consistently.
Without such an expansion parameter, one cannot determine the size of omitted
contributions,
and one can end up retaining certain graphs  while dropping operators
needed to renormalize those graphs.  This results
in a model-dependent description of the short distance physics,
as opposed to a proper effective field theory calculation.

Since the sizes
of the contact interactions depend on the renormalization scheme one uses, the
task becomes one of identifying the appropriate subtraction scheme that makes
the
power counting simple and manifest.
The $\ms$ scheme fails on this point;  however  this is not a
problem with dimensional regularization, but rather a problem with the
minimal subtraction scheme itself.  The momentum space subtraction at threshold
used in
Ref.~\cite{Weinberg}\ behaves similarly.

Next, consider an alternative regularization and renormalization
scheme, namely to using a momentum cutoff equal to $\Lambda$.
Then for large $a$ one finds $C_0 \sim (4\pi/M\Lambda)$, and each additional
loop contributes
a factor of  $C_0(\Lambda + ip)M/4\pi \sim (1+ip/\Lambda)$.
The problem with this scheme is that for $\Lambda\gg p$ the term $ip/\Lambda$
from the loop is small relative to the $1$,
and ought to be ignorable;
however, neglecting it would fail to reproduce the desired result \eq{ami}.
This scheme suffers from
significant cancellations between terms, and so once again the power counting
is not manifest.

Evidently, since $\CA_{-1}$ scales as $1/p$, the desired expansion
would have each individual graph contributing to $\CA_{-1}$ scale as $1/p$.
As the tree level contribution is $C_0$, I must therefore  have
$C_0$ be of size $\propto 1/p$, and each additional loop must be $O(1)$.
This can be achieved by using  dimensional regularization and
 the \pds (power divergence  subtraction)
scheme introduced in  Ref.~\cite{KSW}.
The \pds scheme involves subtracting from the
dimensionally regulated loop integrals not only the $1/(D-4)$ poles
corresponding
to log divergences, as in $\ms$, but also
poles in lower dimension which correspond to power law divergences at $D=4$.
The integral $I_n$ in
 \eq{loopi}\ has a pole in $D=3$ dimensions which can be removed by adding to
$I_n$
the counterterm
\beq
\delta I_n = -{M(ME)^n \mu\over 4\pi (D-3)},
\eeq
so that the subtracted integral in $D=4$ dimensions is
\beq
I_n^{PDS} = I_n + \delta I_n = - (ME)^n \left({M\over 4\pi}\right) (\mu + ip).
\eqn{ipds}
\eeq
In this subtraction scheme
\begin{eqnarray}
\CA & = & -{M\over 4\pi}\[
{4\pi\over M  \sum C_{2n} p^{2n} } + \mu+ip\]^{-1}
\ .
\eqn{answer}
\end{eqnarray}
By performing a Taylor expansion of the denominator of the above expression,
and comparing with \eq{aexp2}, one finds that
for $\mu\gg 1/|a|$,
the couplings $C_{2n}(\mu)$ scale as
\beq
C_{2n}(\mu) \sim {4\pi \over M \Lambda^n \mu^{n+1}}\ .
\eqn{cscale}
\eeq
Eq.~\ref{eq:cscale} implies that  $\mu \sim p$, $C_{2n}(\mu)\sim 1/p^{n+1}$.
A factor of $\nabla^{2n}$ at a vertex scales as $p^{2n}$, while each loop
contributes a factor of
$p$. The power counting rules for the case of large scattering length are
therefore:
\begin{enumerate}
\item{Each propagator counts as $1/p^2$;}
\item{Each loop integration $\int {\rm d}^4q$ counts as $p^5$;}
\item{Each vertex $C_{2n}\nabla^{2n}$ contributes $p^{n-1}$.}
\end{enumerate}
We see that this scheme avoids the problems encountered with the choices of the
$\ms$ ($\mu=0$) or momentum
cuttoff ($\mu\sim \Lambda$) schemes.
First of all, a tree level diagram with a $C_0$ vertex is $O(p^{-1})$, while
each loop with a $C_0$ vertex contributes $C_0(\mu) M (\mu+ip)/4\pi\sim 1$.
Therefore  each term in the bubble sum contributing
to $\CA_{-1}$ is  of order $p^{-1}$, unlike the case for $\mu=0$.  Secondly,
since $\mu\sim p$,
it makes sense keeping  both the $\mu$ and the $ip$ in \eq{ipds} as they are of
similar size, unlike
what we found in the $\mu=\Lambda$ case.
The \pds scheme  retains the nice feature of $\ms$ that
powers of $q$ inside the loop integration are effectively replaced by powers
of the external momentum $p$  \footnote{An alternative subtraction scheme with
similar power counting  is to perform
a momentum subtraction at $p^2=-\mu^2$, as recently suggested in
Ref.~\cite{Gegelia}.}.

\begin{figure}[t]
\centerline{\epsfysize=3 in \epsfbox{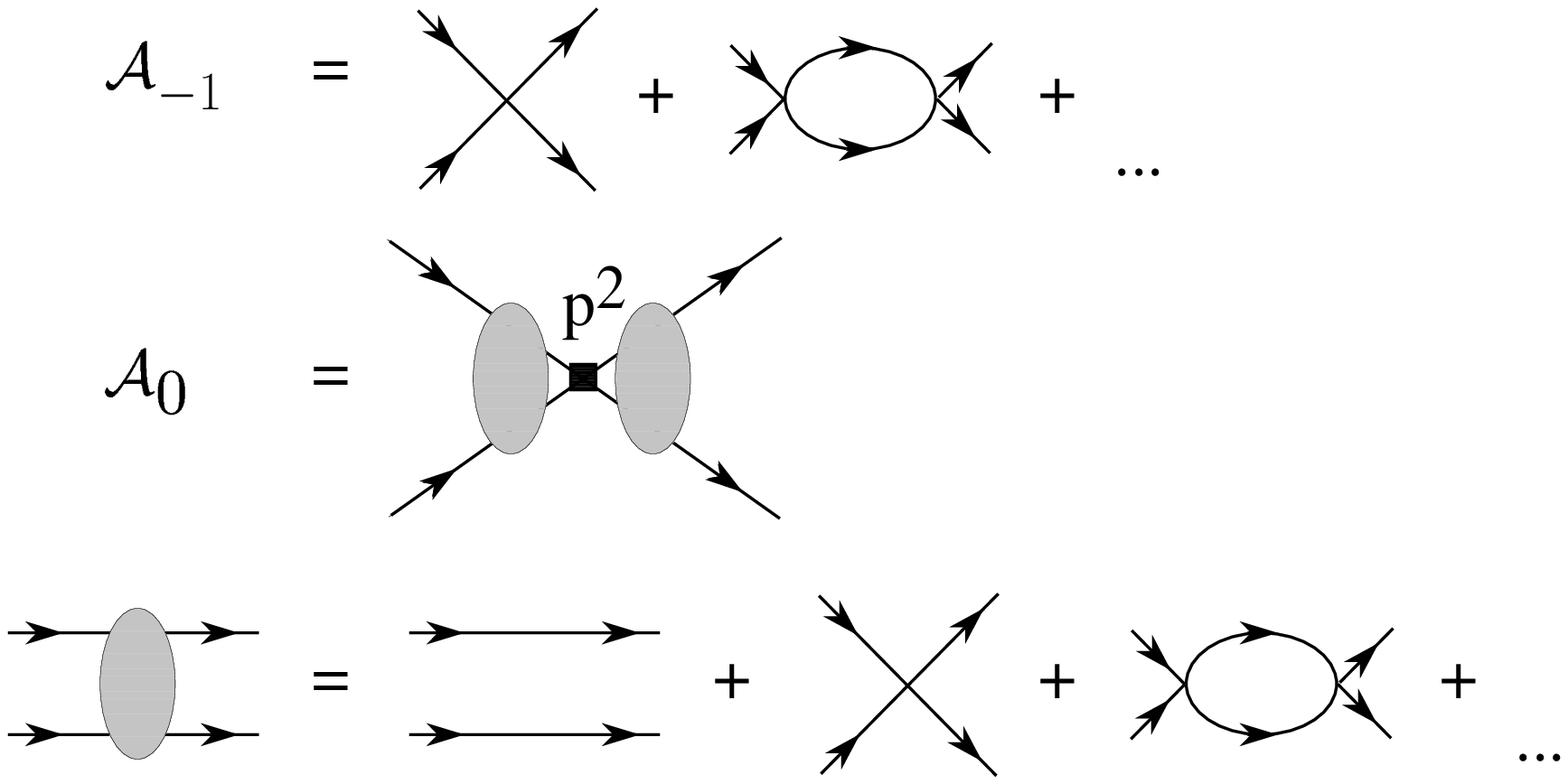}}
\noindent
\caption{\it Leading and subleading contributions arising from local
operators.}
\label{FG1S0_m1}
\vskip .2in
\end{figure}

Starting from the above counting rules one finds that the leading order
contribution to the scattering amplitude $\CA_{-1}$
scales as $p^{-1}$ and consists of the sum of bubble diagrams with $C_0$
vertices;
contributions to the amplitude scaling as higher powers of $p$ come from
perturbative
insertions of derivative interactions, dressed to all orders by $C_0$.  The
first three
terms in the expansion are
\beq
{\cal A}_{-1}&=& { -C_0\over \left[1 + {C_0 M\over 4\pi} (\mu + ip)\right]}\
,\nonumber\\
{\cal A}_0    &=& { -C_2 p^2\over \[1 + {C_0 M\over 4\pi}(\mu + ip)\]^2}\
,\nonumber\\
{\cal A}_1    &=& \({ (C_2 p^2)^2M(\mu+ ip)/4\pi\over \[1 + {C_0 M\over
4\pi}(\mu + ip)\]^3}
-{ C_4 p^4\over \[1 + {C_0 M\over 4\pi}(\mu + ip)\]^2}\) \ ,
\eqn{athree}
\eeq
where the first two correspond to the Feynman diagrams
in Fig.~\ref{FG1S0_m1}.  The third term, $\CA_1$, comes from  graphs with
either one insertion of $C_4\nabla^4$ or two insertions of $C_2\nabla^2$,
dressed to all orders by the $C_0$ interaction.

Comparing \eq{athree} with the expansion of the amplitude \eq{aexp2},  the
couplings $C_{2n}$ are related
to the low energy scattering data $a$, $r_n$:
\beq
C_0(\mu) &=& {4\pi\over M }\({1\over -\mu+1/a}\)\ ,\nonumber\\
C_2(\mu) &=&  {4\pi\over M }\({1\over -\mu+1/a}\)^2 {{
r}_0\over 2}\ ,\nonumber\\
C_4(\mu) &=&  {4\pi\over M }\({1\over -\mu+1/a}\)^3 \[{1\over 4}
{ r}_0^2 + {1\over 2} {{ r}_1\over\Lambda^2} \({-\mu+1/a}\)\]\ .
\eqn{cvals}
\eeq
Note that assuming $r_n\sim 1/\Lambda$, these expressions are
consistent with the scaling law in \eq{cscale}.

%%%%%%%%%%%%%%%%%%%%%%%%%%%%%%%%%%%%%%%%%%%%%%%%%%%%%%%%%%%%
\subsection{The renormalization group}
\label{sec:2c}
%%%%%%%%%%%%%%%%%%%%%%%%%%%%%%%%%%%%%%%%%%%%%%%%%%%%%%%%%%%%

This power counting described in the previous section relies entirely on the
running  of
$C_{2n}(\mu)$ as a function of $\mu$
given in \eq{cscale}.  This was derived by summing up all of the diagrams in
Fig.~\ref{bubbles} explicitly, and then comparing with the form of a general
amplitude, \eq{amp}.   When pions are included in real $NN$ interactions, the
diagrams in Fig.~\ref{bubbles} cannot be explicitly summed.  However, the power
counting can be established perturbatively by examining the $\beta$-functions
and the renormalization group running of the $C_{2n}(\mu)$ couplings.  The
dependence of $C_{2n}(\mu)$ on
$\mu$ is determined by the requirement that the amplitude be independent of the
arbitrary parameter $\mu$.  The physical parameters $a$, $ r_n$ enter as
boundary conditions on the RG equations.

The $\beta$-function $\beta_{2n}$ for the coupling $C_{2n}$ is defined by
\beq
\beta_{2n} \equiv \mu {{\rm d}C_{2n}\over {\rm d}\mu}\ ,
\eqn{betadef}
\eeq
and all of the $\beta$-functions can be computed by requiring that any physical
quantity (e.g. the
scattering amplitude) be independent of $\mu$.
In the \pds scheme, the $\mu$ dependence of the $C_{2n}$ coefficients enters
either logarithmically or linearly, associated with simple $1/(D-4)$ or
$1/(D-3)$ poles respectively.
The functions $\beta_{2n}$  follow straightforwardly from
$\mu{d\over d\mu}(1/\CA)=0$, using the expression for $\CA$
in \eq{answer}.  This gives
\beq
\beta_{2n} = {M\mu\over 4\pi} \sum_{m=0}^{n} C_{2m} C_{2(n-m)}\ .
\eqn{beta2n}
\eeq
\begin{figure}[t]
\centerline{\epsfysize=3 in \epsfbox{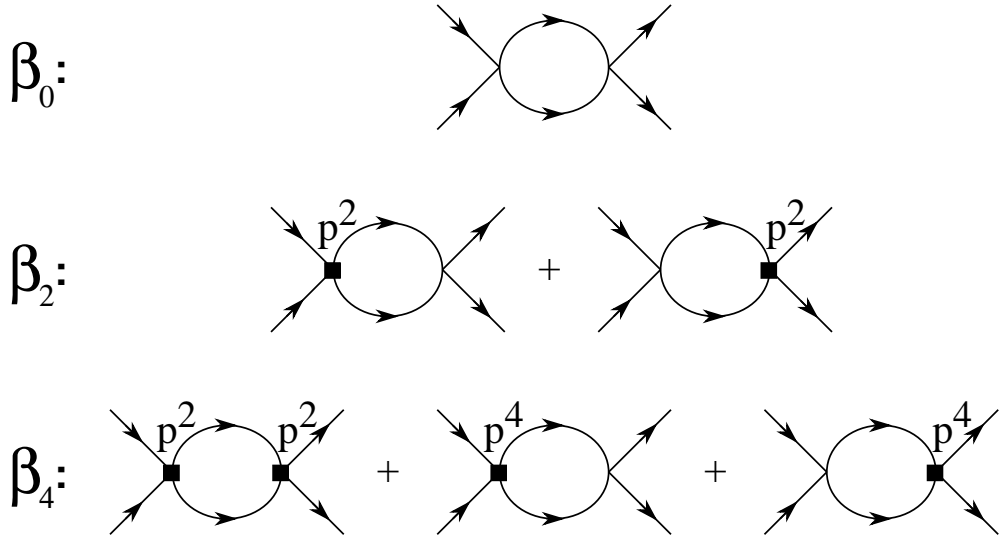}}
\noindent
\caption{\it Graphs contributing to the $\beta$-functions for $C_{2n}$}
\vskip .2in
\label{betanopi}
\end{figure}
However, I did not need the full, explicit amplitude $\CA$ to compute
$\beta_{2n}$, as the exact $\beta$-functions can be computed from the one-loop
diagrams  shown in Fig.~\ref{betanopi}.  That is because only the one-loop
diagrams contribute to simple poles.  Without pions, the only poles one
encounters  are all of the form $1/(D-3)$.

We examine the RG equations for the first two couplings, $C_0$ and $C_2$,
in order to
explicitly show how one recovers the results in \eq{cvals} from solving the
renormalization group equations. From \eq{beta2n}---or equivalently, the
diagrams in Fig.~\ref{betanopi}--- it follows that
\beq
\beta_0&=& {M\mu\over 4\pi} C_0^2\ ,\nonumber\\
\beta_2&=& 2 {M\mu\over 4\pi} C_0 C_2\ .
\eqn{beta024}
\eeq
Integrating these equations relates the $C_{2n}$ coefficients at two different
renormalization scales $\mu$ and $\mu_0$. Comparing the theory with $\CA$ and
its
derivatives at $\mu=p=0$ determines the initial values $C_{2n}(0)$ as in
\eq{cfit}.
The solutions to \eq{beta024} are
\beq
 C_0(\mu) &=& {C_0(\mu_0)\over 1+ C_0(\mu_0)M  (\mu_0 - \mu)/4\pi}\
,\nonumber\\
 C_2(\mu) &=& C_2(\mu_0)\({C_0(\mu)\over C_0(\mu_0)}\)^2\ .
\eqn{c0rg}
\eeq
The boundary conditions are supplied by equating the computed $\CA$ to
scattering data at $\mu=p=0$ \footnote{This is not the only choice.  For
example, when discussing the deuteron, it is convenient to fix the location of
the pole to the true deuteron binding energy.  The differences arising from
this alternate fitting procedure are small higher order effects.}, which yields
\beq
C_0(0)&=&4\pi a/M\ ,\qquad  C_2(0) = C_0(0) a r_0/ 2\ ,
\eeq
and so forth.
 With these boundary conditions and the solutions \eq{c0rg}
we arrive at the result derived previously for $C_0(\mu)$ and  $C_2(\mu)$ in
\eq{cvals}:
\beq
C_0(\mu) = {4\pi\over M }\({1\over -\mu+1/a}\)\ ,\nonumber\\
C_2(\mu) = {4\pi\over M }\({1\over- \mu+1/a}\)^2 {{
r}_0\over 2}\ .
\eqn{c0run}
\eeq

It is instructive to solve the complete, coupled RG equation
\beq
\mu {{\rm d}\ \over {\rm d}\mu}C_{2n} = {M\mu\over 4\pi}\sum_{m=0}^n C_{2m}
C_{2(n-m)}
\eeq
for the leading small $\mu$ behavior of each of the coefficients $C_{2n}$ . The
solution, for $n>0$ is
\beq
C_{2n}(\mu) = {4\pi \over M(-\mu+1/a)} \({r_0/2\over -\mu+1/a}\)^n +
O(\mu^{-n})\ .
\eeq
First note that the scaling property in \eq{cscale} is realized:
$C_{2n}(\mu)\propto \mu^{-(n+1)}$ for
$|1/a|\ll\mu\ll\Lambda$.  What is curious is that this leading behavior does
not entail a new
integration constant for each $n$, but only depends on the two parameters $a$
and $r_0$
encountered when solving for $C_0(\mu)$ and $C_2(\mu)$;  this is due to a
quasi-fixed
point behavior of the RG equations ---  the $C_{2n}$ couplings are being driven
primarily
by lower dimensional interactions. One can see this explicitly in our formula
\eq{cvals}
for $C_4$, where the leading $O(\mu^{-3})$ part of $C_4$ depends only on $r_0$,
while
the subleading $O(\mu^{-2})$ part is proportional to $r_1$.

This behavior allows us to establish a connection between the present work, and
the
method of introducing an $s$-channel dibaryon discussed in 
Ref.~\cite{DBK}.  The leading
$\mu$ behavior of all of the $C_{2n}$ coefficients is determined by the
effective range $r_0$.
If one resums this leading behavior at the $\tilde N \tilde N$ vertex one finds
(for
$\mu\sim p\gg 1/|a|$)
\beq
\sum_{n=0}^\infty C_{2n}(\mu) p^{2n} &=&
{4\pi \over M}{1\over -\mu+1/a -{r_0 p^2\over 2}}+ O(p^2)\nonumber\\
&=&-{(8\pi /  M^2 r_0)\over E-(-\mu+1/a)/Mr_0}\ .
\eeq
This looks like an $s$-channel propagator for a particle at rest of mass
$[2M + (-\mu+1/a)/Mr_0]$, and in fact, for $\mu=0$, corresponds exactly to the
dibaryon proposed in Ref.~\cite{DBK} to reproduce the scattering
due to a short range potential.  We see that using the dibaryon is as good as
(but no better than)
 carrying out the effective field theory calculation to $O(p^1)$. The
subleading
corrections can be accounted for by including the subleading part of the $
C_4(\mu)
p^4$ vertex proportional to $r_1$, and which occurs at  $O(p^2)$.   This
dibaryon
was recently used with great success in the three-body problem~\cite{Bvk}.

%%%%%%%%%%%%%%%%%%%%%%%%%%%%%%%%%%%%%%%%%%%%%%%%%%%%%%%%%%%%
\section{Expanding the potential or the scattering amplitude?}
\label{sec:3}
%%%%%%%%%%%%%%%%%%%%%%%%%%%%%%%%%%%%%%%%%%%%%%%%%%%%%%%%%%%%

Throughout this talk I have discussed an expansion of the scattering amplitude,
while most previous work in the subject of effective field theory for $NN$
scattering has emphasized expansion of the
potential\cite{Weinberg,Bira,Park,KSWa,CoKoM,Cohen,Lepage,Adhik,RBMa,Gegelia,Steele},
followed by a solution of the Lippmann-Schwinger equation with this 
approximate potential.  Since every observable is an $S$-matrix element, it 
would seem that the two  methods, if carried out to the same order, ought to 
give answers that disagree only by higher order effects.  For example, if
 working to $O(p^0)$, the direct expansion of $\CA\simeq \CA_{-1}+\CA_0$ is 
linear in the $C_2\nabla^2$ operator.  Alternatively, if the  potential is 
expanded to linear order in $C_2\nabla^2$, the Lippmann-Schwinger equation 
yields an amplitude which reproduces the term linear in $C_2$, but also includes
 terms higher order in $C_2$.  These terms nonlinear in $C_2$ would appear to 
be higher order effects  and negligible.

  This is not in general true, however.  Unlike the $\CA$ expansion I have
outlined which is explicitly  scheme independent, the Lippmann-Schwinger
approach is not.  That is because the time ordered product of several
insertions of the potential induces new divergences that require counterterms
not included in the expansion.  For example, when the potential is taken to
include only the $C_0$ and $C_2\nabla^2$ operators, the one-loop contribution
to the Lippmann-Schwinger equation has a divergence that requires the
$C_4\nabla^4$ operator to absorb the divergence---an  operator  not  included
in the expansion!  As a result,  the Lippmann-Schwinger approach is necessarily
scheme-dependent.

Is this bad?  Not fatal, but undesirable.  It means that the size of neglected
effects depends  on the renormalization scheme and cutoff or renormalization
scale.  That is why in calculations of this sort, the renormalization scale or
cutoff becomes an extra parameter that has to be chosen to minimize the errors.

Understanding the scheme dependence of the Lippmann-Schwinger result helps
explain an apparent paradox:  the amplitude $\CA$ calculated in this talk in
the $\pds$ scheme is $\mu$ independent at each order in the expansion.  In
particular, each term $\CA_n$ is unchanged if I take $\mu\to 0$, which is the
$\ms$ scheme.  However, there have been numerous discussions about how $\ms$ is
sick.  In fact, $\ms$ is not sick except for the fact that at any given order
there tend to be large cancellations among the graphs included at that order.
However, in the Lippmann-Schwinger approach, {\it parts} of higher order terms
are kept, which destroys the cancellations; the result is $\mu$ dependent, and
the expansion becomes very bad for $\mu\ll p$.  For example, at $O(p^1)$,
$\CA_1$ given in \eq{athree} involves both a $C_2^2$ term and a $C_4$ term; for
$\mu\ll p$ these two terms are both large and cancel against each other.
However the $O(p^0)$ Lippmann-Schwinger calculation, in which the potential
includes the $C_2$ interaction, but not the $C_4$ term, gives a scattering
amplitude that includes the $C_2^2$ piece of $\CA_1$ but not the $C_4$ piece
which is needed to largely cancel the $C_2^2$ contribution and render its
effects small. If one solves the Lippmann-Schwinger equation {\it and} chooses
the $\pds$ value $\mu\sim p$, then the $C_2^2$ and $C_4$ terms in $\CA_1$ are
of the same size and (by construction) and small, and the error in the
Lippmann-Schwinger amplitude really is higher order.  This has been
independently discussed in Ref.~\cite{Gegelia}.

The correct statement about the $\ms$ subtraction scheme is that it can be
misleading, in that one cannot look at an individual Feynman graph at
contributing to $\CA_n$ and expect it to contribute at $O(p^n)$. The problems
with the $\ms$ scheme reported in  the
literature~\cite{KSWa,Cohen,Lepage,Steele} are the result of combining the
$\ms$ scheme with the scheme dependent Lippmann-Schwinger approach.  I favor
avoiding the Lippmann-Schwinger approach altogether---why would one want a
calculational scheme where the size of the errors at a given order in the
expansion depend on the renormalization scheme?  In effective field theory,
with its necessarily singular interactions that require renormalization, the
concept of a classical potential does not seem very useful.

%%%%%%%%%%%%%%%%%%%%%%%%%%%%%%%%%%%%%%%%%%%%%%%%%%%%%%%%%%%%
\section{Conclusions and a challenge}
\label{sec:4}
%%%%%%%%%%%%%%%%%%%%%%%%%%%%%%%%%%%%%%%%%%%%%%%%%%%%%%%%%%%%

I hope I have convinced you that there exists a rather simple and well defined
approach to calculating low energy processes involving two nucleons.  There are
quite a few interactions  of interest to calculate, such as $NN\to NN\gamma$,
$NN\to d\gamma$, $pp\to d e^+\nu$, $\pi d\to \pi d$, as well as isospin
violation and parity violation in $NN$ scattering.  And for every calculation,
there is always one higher order that can be computed!

However, it would be disappointing if the applicability of these techniques
were limited to two nucleon processes. Evidence to date~\cite{KSW,Deut}
suggests that the expansion works reasonably well up to momenta comparable to
the Fermi momentum in nuclear matter.  What are the prospects of using
effective field theory to discuss the structure of nuclei, or the equation of
state of nuclear matter?

At present the road block is the three-body interaction.  It is tempting to
dismiss the three-body interaction as negligible, as often claimed.  However,
the two-body interactions renormalize the three-body force, which means that in
fact the strength of the three-body interaction is scheme dependent, and saying
it is small in general is nonsensical.  In fact, because of the
Efimov~\cite{Efimov} and Thomas~\cite{Thomas} effects, we know that we cannot
keep zero-range two-body interactions while neglecting  three-body
interactions.  It is clear, due to Pauli statistics, that at most there can be
four-body contact interactions (without derivatives).  However, until the
properties of the three- and four- body interactions are understood properly,
it is unclear how to proceed to a discussion of nuclear matter.  I view
understanding this issue of few-body interactions to be the fundamental
challenge in this field.

%%%%%%%%%%%%%%%%%%%%%%%%%%%%%%%%%%%%%%%%%%%%%%%%%%%%%%%%%%%%
\section*{Acknowledgments}
%%%%%%%%%%%%%%%%%%%%%%%%%%%%%%%%%%%%%%%%%%%%%%%%%%%%%%%%%%%%

I would like to thank the organizers of this joint workshop between the Kellogg
Radiation Lab and the Institute for Nuclear Theory---U. van Kolck, M. Savage,
and R. Seki---for putting together such an enjoyable and stimulating meeting. I
would also like to thank R. McKeown for helping make this happen. This work
supported in part by the U.S. Dept. of Energy under Grant No. DOE-ER-40561.


\section*{References}
\begin{thebibliography}{99}
%
\bibitem{Martin} M. J. Savage, {\it Including Pions}, {\tt nucl-th/9804034}.
%
%
\bibitem{KSW} D.B. Kaplan, M.J. Savage and M.B. Wise,
{\tt nucl-th/9801034}, {\it to appear in Phys. Lett. B};
{\tt nucl-th/9802075}, {\it submitted to Nucl. Phys. B}.
%
%
\bibitem{Weinberg}S. Weinberg,
Phys. Lett. {\bf B251} (1990) 288;
Nucl. Phys. {\bf B363} (1991) 3;
Phys. Lett. {\bf B295} (1992) 114.
%
%
\bibitem{Deut} D. B. Kaplan, M. J. Savage, M. B. Wise, {\tt nucl-th/9804032},
{\it submitted to Nucl. Phys. B}
%
%
\bibitem{Bira} C. Ordonez and U. van Kolck, Phys. Lett. {\bf B291} (1992) 459;
C. Ordonez, L. Ray and  U. van Kolck, Phys. Rev. Lett. {\bf 72} (1994) 1982;
Phys. Rev. {\bf C53} (1996) 2086.;
U. van Kolck, Phys. Rev. {\bf C49} (1994) 2932.
%
%
\bibitem{Park} T.-S.  Park, D-.P.  Min and M. Rho,
Phys. Rev. Lett. {\bf 74} (1995) 4153;
Nucl. Phys. {\bf A596} (1996) 515;
 T.-S. Park, K. Kubodera, D.-P. Min, M. Rho,
{\tt  hep-ph/9711463}. %
%
%
\bibitem{KSWa} D.B. Kaplan, M.J. Savage and M.B. Wise,
\Journal{\NPB}{478}{629}{1996},
{\tt nucl-th/9605002}.
%
%
\bibitem{CoKoM} T. Cohen, J.L. Friar, G.A. Miller and
U. van Kolck,
\Journal{\PRC}{53}{2661}{1996}.
%
%
\bibitem{Cohen}T.D. Cohen,
\Journal{\PRC}{55}{67}{1997}.
D.R. Phillips and T.D. Cohen,
\Journal{\PLB}{390}{7}{1997}.
K.A. Scaldeferri, D.R. Phillips, C.W. Kao and T.D. Cohen,
\Journal{\PRC}{56}{679}{1997}.
S.R. Beane, T.D. Cohen and D.R. Phillips,
nucl-th/9709062.
%
%
\bibitem{Lepage} G. P. Lepage, {\it How to renormalize the Schr\"odinger
Equation},
 Lectures given at 9th Jorge Andre Swieca Summer School: Particles and Fields,
Sao Paulo, Brazil, 16-28 Feb 1997.
{\tt nucl-th/9706029}.
%
%
\bibitem{Adhik} S.K. Adhikari and A. Ghosh,
J. Phys. {\bf A30}, 6553 (1997).
%
%
\bibitem{RBMa}  K.G. Richardson, M.C. Birse and J.A. McGovern,
{\tt hep-ph/9708435}.
%
%
\bibitem{Gegelia} J. Gegelia,
{\tt  nucl-th/9802038}.
%
%
\bibitem{Steele} J.V. Steele and R.J. Furnstahl,
{\tt nucl-th/9802069}.
%
%
\bibitem{ManRev} A.V. Manohar,
Lectures given at 35th Internationale Universitatswochen f\"ur Kern- und
Teilchenphysik,
{\it Perturbative and Nonperturbative Aspects of Quantum Field Theory},
Schladming, Austria, 2-9 Mar 1996.
{\tt hep-ph/9606222 }.
%
%
\bibitem{DBK} D. B. Kaplan, Nucl. Phys. {\bf B 494} (1997) 471, {\tt
nucl-th/9610052}.
%
%
\bibitem{Bvk} P.F. Bedaque and U. van Kolck,
{\tt nucl-th/9710073};
P.F. Bedaque, H.-W. Hammer and U. van Kolck,
{\tt nucl-th/9802057}.
%
%
\bibitem{Efimov} V. N. Efimov, Sov. Jour. of Nucl. Phys. {\bf 12} (1971) 589.
%
%
\bibitem{Thomas} L. H. Thomas, Phys. Rev. {\bf 47} (1935) 903.
\end{thebibliography}
\end{document}